\documentclass[sigconf]{acmart}
\AtBeginDocument{%
  }
\usepackage{multirow}
\usepackage{placeins}
\usepackage{hyperref}
\hypersetup{colorlinks,allcolors=black}

\acmConference{Defective}{Amplified}{Contested}
\setcopyright{none}
\renewcommand\footnotetextcopyrightpermission[1]{}

\begin{document}

\title{Defective, Amplified, Contested:\\ Folk Conceptualizations of AI Slop}

\author{Zhixin Li}
\email{zli457@syr.edu}
\orcid{0009-0006-1118-4361}
\affiliation{%
\department{School of Information Studies}
  \institution{Syracuse University}
  \city{Syracuse}
  \state{NY}
  \country{USA}
}

\author{Jaime Banks}
\email{banks@syr.edu}
\orcid{0000-0002-7598-4337}
\affiliation{%
\department{School of Information Studies}
  \institution{Syracuse University}
  \city{Syracuse}
  \state{NY}
  \country{USA}
}

\renewcommand{\shortauthors}{Li and Banks}

\begin{abstract}
AI slop has flooded social media feeds with overprocessed and derivative content—Jesus fusing with a shrimp, a humanoid orange chubby cat shoplifting, and a “hero” dog rescuing a human child after a plane crash. This might be no laughing matter, as waves of media coverage have taken a distinctly negative bent, emphasizing its low quality, ethical concerns, and threats to human creativity. Despite extensive attention from politicians and technologists, the perspectives of everyday users—the very audience most exposed to this phenomenon—remain largely unexplored. This descriptive work explored how Reddit users experience, conceptualize and contest AI slop. Inductive thematic analysis of posts and comments across six slop-related subreddits identified four themes covering constitutional defects, algorithmic overreach, user resistance, and entertainment outcomes—ranging from anxieties about eroding agency, moral norms, creativity, and existential questions to defensive distinctions between quality AI content and slop. We derive a user-driven definition: \textit{AI slop is audience-contested AI-generated content characterized by constitutional defects and enabled by platform logics}. We operationalize slop through a facet-theoretical framework to inform future research into this fuzzy category.
\end{abstract}



\maketitle
\pagestyle{fancy}
\fancyhf{}                                   
\fancyhead[L]{\small Conceptualizing AI Slop}
\fancyhead[R]{\small Li and Banks}
\fancyfoot[C]{\thepage}                      
\renewcommand{\headrulewidth}{0pt}       

\section{Introduction}
One 60-second video making its rounds features a humanoid, red-haired cat in ragged clothes, furious as poverty has worn her down—she’s had enough. She hurls the plate at a male ginger cat cooking in their rundown kitchen. But love persists. He works harder, saving every coin for a ring. Then betrayal comes swiftly: A wealthy white cat appears, driving in a luxury car with an oddly rendered third leg, buying expensive dresses at high-end boutiques. She’s gone—whisked away by a white cat. The ginger cat’s world collapses, the video ending with his incongruously human fingers throwing their framed photo onto a dusty floor. This AI-generated saga has amassed 186.5 million views. Such digital kitsch has earned a derisive label: AI “slop,” which has come to describe low-effort, mass-produced, and derivative AI-generated content (e.g., texts, images, and videos) that saturates social media feeds. The term often carries a negative connotation, suggesting content that is inaccurate, deceptive, unreliable, and cluttering \cite{Madsen2025b}. Similar framings guide scholarly and popular discourse, but realities about how slop is created, spread, encountered, consumed, and contested are more complex.

\subsection{The Emergence of Slop}
The term “slop” long predates generative AI, tracing to around 1400, from Old English “-sloppe” (dung) and “slyppe” (slime), denoting waste and low-quality material, later extended to mean animal feed and refuse \cite{OnlineEtymology}. “Slop” circulated online throughout the 2010s, particularly on anonymous forums, such as 4chan, as a pejorative shorthand for cultural content (e.g., movies) perceived as low-quality, homogenized, and lazily produced\footnote{We searched 4chan archives (4plebs) for earlier uses of the term and found no instances before 2010; the earliest post we located dates to February 2010, suggesting the term mainstreamed around that time.}. From that vernacular, the label became reoriented toward AI-generated images following the release of text-to-image models such as DALLE 2, Midjourney, and Stable Diffusion, tools widely perceived as potential threats to human creativity \cite{Banks2025}. On June 24, 2025, AI slop was officially recognized when Cambridge Dictionary \cite{CambridgeDictionary2025} added a new AI-related definition of slop: “content on the internet that is of very low quality, especially when it is created by artificial intelligence.” 

Operationalizing the term, AI slop has spread across media formats, with each evolution amplifying its capacity for viral engagement. In text form, hallucinations proliferated following the 2022 public rollout of large language models, ranging from entirely fictional academic references to invented historical facts. In images, viral AI-generated content ranged from a photorealistic Pope Francis in a puffer jacket to images of political figures in fabricated scenarios. In audio, slop soon cluttered playlists, prompting Spotify to remove 75 million “spammy” AI-generated tracks amid complaints from artists and users \cite{Robinson2025}. AI slop then expanded into videos, where now-defunct Sora 2 examples dominated social media with hyper-realistic AI clips.

\subsection{Slop in the Wild}

AI slop (henceforth “slop”) has become a fashionable topic in media discourse, with news headlines tending to surge around sensational or ethically charged events. For instance, Sora 2 digitally revived deceased public figures such as Robin Williams and Martin Luther King, Jr. in AI-generated videos, moving their families to fight for their dignity \cite{Sarwar2025SoraDeadCelebs}; the billionaire YouTube star MrBeast has expressed fears that AI-generated replicas could undermine his livelihood \cite{Wilkins2025MrBeast}; and AI-generated imagery has even been used to promote dubious health supplements \cite{Given2025GenerativeAI}. 

The major social media platforms—TikTok, YouTube, Facebook, and Instagram—however, have become significant facilitators of slop by actively fueling and funding slop through creator monetization programs while simultaneously developing their own AI content creation tools. For instance, Google DeepMind’s Veo 3 was integrated into YouTube Shorts to make content creation even more “effortless” \cite{Berrada2025YouTubeAI}. TikTok’s Creator Fund Program was expanded to Creator Rewards Program for content creation mainly based on virality (i.e., 100,000 + views in the past 30 days as of August 2026), making AI-generated viral content economically attractive. Instagram opened the floodgates for far-less-expensive AI influencer accounts who function alongside human influencers, endorsing partnered brands that generate millions of revenues. Against this backdrop, news coverage have converged around blame and responsibility: OpenAI Sora 2 is considered at-fault in copyright infringement \cite{MPA2025SoraStatement}, Meta Vibes is blamed for lowering feed quality \cite{Stanton2025MetaVibes}, and even LinkedIn, which Reddit cofounder Alexis Ohanian pointed to as an example of the internet being “quasi-AI” and increasingly “dead” \cite{Chandonnet2025Ohanian}.

In response to expert- or society-level pushbacks, social media platforms have responded unevenly, ranging from tentative content-moderation efforts to outright defenses of AI-generated content as a natural evolution of their platforms. Pinterest is stemming the tide of slop with new GenAI filtering tools \cite{Pinterest2025GenAIControl}; TikTok is partnering with the Coalition for Content Provenance and Authenticity (C2PA) and expanding auto-labeling to AI-generated content. YouTube CEO Neal Mohan announced “managing AI slop” is a priority for 2026 \cite{Mohan2026FutureYouTube}. Yet Meta is dismissive and defensive on the matter. Connor Hayes, vice president of product for generative AI at Meta, expected those AI accounts “to actually exist on our platforms in the same way that [human] accounts do” \cite{Criddle2024MetaAIUsers}. Similarly, in a recent interview with Computer Network, Facebook’s vice president of product, Jagjit Chawla, proclaimed that slop is algorithm-sensitive and Facebook treats it the same as human-generated content \cite{Chedraoui2025FacebookReels}.

This algorithmic approach places the burden of content curation squarely on users: whenever they send “positive” signals (e.g., likes, shares) to AI-generated content on their feeds, the recommendations algorithm will determine that users are, in fact, interested in slop. This approach assumes signals are authentic, though concerns have been raised about coordinated bot activity artificially inflating engagement with social media content \cite{Molina2022}. To reduce unwanted slop content, platforms suggest users employ the “Not interested” button, yet research reveals these disinterest tactics or algorithmic resistance may often seem ineffective \cite{Vera2025}. In this sense, users become rather passive as they cannot like it, comment on it, or share it.

Yet users may encounter a different set of incentives when they occupy the creator role.  Participatory culture lowers barriers to creative expression and enables users to take a more active role in producing and circulating media \cite{Jenkins2009Confronting}, while generative technologies further reduce the cost and effort of production. Content users despise as tasteless slop in their feeds may therefore become appealing to produce when it promises  creator-reward payouts or a viral moment \cite{Lee2026}. For creators whose income depends on platform visibility, refusing to participate in these production logics may be not just impractical but economically untenable.

\subsection{Scholarly Characterizations of Slop}

Slop’s virality has grown in parallel with emerging scholarly interest, though much of this scholarship remains conceptual. Stanusch et al. \cite{Stanusch2025} argued the production of synthetic content is “cheap, easy, and fast” and seems to “crack the code for lucrative algorithmic vitality” (p. 6). Similarly, Madsen and Puyt \cite{Madsen2025a} characterized cultural unease by describing slop as “the flood of lazy, derivative, endlessly recycled [content] (p.1).” Madsen and Puyt \cite{Madsen2025b} conceptualized slop variations at the intersection of application domains (e.g., in academia, politics, media) and big-data properties: Volume (expansive production scales), Velocity (high speed of generation and circulation), Variety (wide range of forms and genres), Value (erosion of cultural and epistemic worth), Verification (problems of truth and trust), Visibility (algorithmic amplification), and Virality (meme-like novelty and diffusion). Two empirical studies have provided operationalizations of slop. Shaib et al. \cite{Shaib2025} developed a taxonomy of slop and proposed interpretable dimensions—information utility, information quality, style quality—for its assessments in text, finding slop judgements are not binary. Stanusch and colleagues \cite{Stanusch2025} identified categories akin to genres of AI-generated videos across TikTok and Instagram, including authority satire (e.g., infantilization of authority figures depicted as toddlers), plausible (dis)information (e.g., seeming amateur footage of natural disasters), illusive historicizing (e.g., distortions of historical events), and surrogate clips (e.g., real figures shown in historically impossible contexts or actions). These conceptual and empirical developments together frame slop as a multidimensional phenomenon characterized by low informational utility, questionable quality, and formulaic style. 

These characterizations begin to transform “slop” from a loose label into an analyzable phenomenon. The remaining question is what kind of phenomenon it is: is it a digital waste, algorithmic content, cultural expression, or some unstable combination of all three? Some argue slop should not be dismissed merely as digital detritus but taken seriously as cultural form with possible social function, aesthetic appeal, and meaning-making potential because it offers a supply-side solution to unmet demand for highly specific, personalized content humans cannot or will not produce at sufficient scale \cite{Kommers2026}. Others counter this perspective risks overlooking the broader platform and sociotechnical conditions shaping slop (e.g., hidden social costs, privacy concerns, and the erosion of creative communities; \cite{Nishal2026}). This debate situates slop as a complex phenomenon shaped not only by judgements of quality but also by broader social consequences still being empirically explored.

\subsection{Considering Folk Understandings of Slop}
In addition to the problematics of slop, we must acknowledge it is part of fast-scrolling platform cultures in ways that may offer personal gratifications. Short-form videos on platforms such as TikTok, Instagram Reels, YouTube Shorts have surged in popularity and now function as a dominant source of gratifications for many users \cite{Buf2020}. As new media technologies evolve and AI is embedded in short-form platforms, rapid, low-effort visual stimulation may generate value for both content creators and consumers. For creators, advanced AI editing tools such as Veo 3, Meta AI, and Kling AI have made the creation of short videos easier and more appealing, satisfying creators’ desires for efficiency, visibility, and self-expression \cite{Hatamleh2025,Kang2022}. With a few taps, creators can transform still images into short video formats using thousands of AI-generated templates, generate visual sequences from a brief prompt, or apply AI voiceovers to endorsed brands instead of spending hours drafting scripts. 

For content consumers, slop can be gratifying through similar mechanisms—it is instantly enjoyable, easily digestible, and caters to reduced attention spans \cite{Chen2023,Nguyen2025}. Human brains are especially responsive to rapid scene changes, vibrant colors, and other high-salience visual stimuli \cite{Stolte2021}, which can make AI video slop difficult to look away from. In this sense, slop may function as a form of digital kitsch with immediate sensory appeal. More generally, it is thought to be an entertaining and otherwise gratifying part of online cultures—much the way early graphic memes were critiqued in broader discourse but enjoyed on individual screens \cite{Miltner2014}. It may be that slop fulfills other gratifications commonly associated with short-form video platforms, including information seeking, social connectedness, escapism, and agency \cite{Vaterlaus2025}. Conversely, slop may be understood to be a villain in the “enshittification” of the internet or more formally “platform decay” \cite{Doctorow2023}, in which online spaces are perceived to be materially degraded (see \cite{Kendall2026}). 

In considering this tension, there is little systematic inquiry into folk conceptualizations--those from everyday, non-expert users like content creators and consumers who most directly encounter the content.  As they may be best positioned to offer grounded understandings, experiences, and opinions the current study takes a step toward bridging that gap by centering public discussions about slop, including commentary about encounters with, reactions to, and sensemaking around slop. We employed an inductive approach to explore and describe the topics addressed in the discussions, to both capture the state of opinion around the phenomenon and to detect signals of useful future research directions. We ask and answer: \\\textbf{RQ}: \textit{How do everyday users conceptualize slop in public discussions?}

\section{Method}

We conducted an inductive, descriptive analysis of posts and comments on slop-focused subreddits, capturing public online discussions on the topic between January 1, 2023 and October 7, 2025. 

\subsection{Sites and Sampling}
To seek out public discourse surrounding slop, Reddit was selected because it provides naturally occurring texts that enable the examination of collective behavior, discourse, and sentiment. Reddit users are also self-organized into topical subreddits, making the platform simultaneously a reflection of and a contributor to broader online culture \cite{LizamaMue2022}. Furthermore, Reddit’s API accessibility and transparent community norms make it a reliable site for large-scale data collection, distinguishing it from platforms with stricter data access or privacy restrictions \cite{Proferes2021}.

Data was collected from six subreddits selected based on two criteria: forums that (a) broadly discuss slop and its social implications (r/AIfreakout, and r/aifails), and (b) focus on dominant content-sharing and visual-centric platforms where slop is frequently encountered (r/TikTokCringe\footnote{r/TikTok was not included because no search results were found as of data collection time.}, r/youtube, and r/instagram, r/FacebookAIslop). The corpus consisted of posts and comments containing the focal term “AI slop” that were posted between January 1, 2023 (shortly after the public release of ChatGPT that mainstreamed public interest in AI-generated content) and October 7, 2025 (a week after the release of Sora 2, OpenAI’s flagship text-to-video generation model).

\subsection{Data Collection and Cleaning}
Data collection was conducted by the first author using the Python Reddit API Wrapper to access Reddit’s API. Posts were included if “AI slop” or variants (“ai-slop,” “aislop,” “ai\_slop”) appeared in the post title, body, or comments. A two-channel collection strategy was employed. First, Reddit’s search endpoint, using CloudSearch query syntax with weekly time segmentation, was used to retrieve posts containing the target phrase in titles or body text. Second, each subreddit’s new feed was automatically scanned to capture posts where the phrase did not appear in the title or body text but appeared in comments. All content was systematically validated using a case-insensitive regular expression \verb|(\bai[\s\_-]?slop\b)| to confirm matches. Posts were deduplicated by subreddit and post ID; comments by subreddit and comment ID. This process resulted in a corpus of \textit{N} = 516 posts and comments. 

\subsection{Analytic Approach}
The data was subjected to inductive analysis to construct a thematic hierarchy of codes, subthemes, and higher-order themes. Given the focus on visual-centric platforms where slop is frequently encountered and discussed, analysis required engagement with both textual and visual elements. The first author reviewed linked media content to understand what posters were responding to and to contextualize their commentary about visual features. This dual engagement—making sense of the visual artifacts while analyzing textual discourse about them—was necessary for accurate interpretation of the text \cite{Trombeta2022}, and coding focused only on how users textually articulated their perceptions. Following the six-step thematic analysis workflow recommended by Braun and Clarke \cite{Braun2006}, the first author: (1) familiarized with data through careful reading, (2) assigned an initial, idiosyncratic code to each chunk of data representing a coherent idea, such that individual posts or comments could be assigned one or more codes depending on the complexity of the expressions, (3) searched for themes by deduplicating those initial codes and iteratively collapsed codes into categories (i.e., subthemes), and then in a subsequent phase collapsed those categories into broader themes, (4) reviewed themes for internal coherence and alignment with the original data, (5) defined and named these themes, and (6) reported the hierarchy with representative data excerpts.

Then, the second author examined the initial thematic hierarchy for face validity, internal consistency, and mutual exclusivity of themes in relation to the source data. After identifying opportunities to collapse discrete categories into richer higher-order categories and split overly broad themes, the thematic hierarchy was reconfigured through discussion between the two authors. 

Analysis was inherently sensitized \cite{Bowen2006} by the first author’s epistemological stance, approaching discussions of slop as multifaceted and socially negotiated practices rather than objective assessments \cite{Burr2015}. This interpretive lens was further shaped by being positioned as a regular consumer of AI-generated content and active member of AI discussion communities, which provided insider knowledge of community vernacular. This positionality facilitated recognition of implicit references (e.g., acronyms, allusions) and in-group terminology (e.g., strawberry diaper cat) while necessitating reflexivity to ground interpretations in posters’ articulated perspectives.

\section{Results}
The inductive analysis resulted in the identification of four themes, each respectively reflecting criteria for, framing of, reactions to, and gratifications from AI slop. Throughout, we include data excerpts to illustrate the induced themes; these excerpts have been slightly altered to prevent direct searchability (and so direct identification of poster user accounts). The thematic hierarchy is presented in Table \ref{tab:slop_thematics}. 

\subsection{Constitutional Defects}
Users consistently grounded their judgments of AI-generated media in the visible deficiencies of the content itself. Across discussions, users framed slop as fundamentally flawed in substantive integrity, coherence, and/or credibility. 

\textbf{Visual degradation} is the most overt and commonly mentioned diagnostic marker, interpreted as tell-tale signs of “sloppiness.” Degradation can be understood as deviations from expected correspondence between media content and real-world referents \cite{Cho2012}. Problematic content can involve merged anatomies, distorted faces, and textual corruption. The descriptions converged primarily around characters whose torsos, limbs, or facial structures do not belong to the same species, such as a humanoid figure rendered with animal musculature and posture. That pattern is exemplified in depictions of cat characters that often feature humanoid bodies that posters recognize as “juxtaposition” of ontologically different features they heuristically “really dislike.”. Other discussions noted scenes featuring multiple distorted and blended faces with oddly cartoonish or uncanny textures and facial features that were visibly misaligned. Instances of misspellings also signaled degraded production quality, with examples like signage reading “expror” instead of “exit” cited as evidence of textual corruption.

Posters also commented on depicted \textbf{moral transgressions}, where users flagged eroded ethical boundaries or amplified socially harmful scenarios involving harm, injustice, or violation of rights \cite{Turiel1983}. A large cluster of discussions featured mistreatment of animals (many of which were often child-coded), such as distressed puppies left on train tracks or “kittens dunked into a blender”. The prevalence of such content prompted posters to question the patterns: “Why are so many AI slop videos about animal abuse?” and sarcastically remark, “Baby wake up, we have a serial child abductor.” Others flagged content involving sexual or gender-based harms, including depictions suggestive of rape, harassment, transphobia, homophobia, or sexist power dynamics. Users reacted with alarm to emerging patterns: “They’re making rape AI slop now,” and “Why are these videos sexualizing the cat people?” Scenes implying murder, arson, theft, or destruction similarly provoked strong condemnation: A seemingly innocent diaper cat goes on a “crime spree”—poisoning other animals, grinding them into meat for money—alongside “scary footage of crazy people destroying bridges with people on them.” Other morally transgressive scenarios included demonizing specific groups, with black cats consistently cast as villains: “Why do they always make the black cat evil?” and “Racist cats are a staple of AI slop.”

\textbf{Narrative inconsistency} was also considered a core defect of video slop, defined as discrepancies or contradictions in logical flow, plot coherence, character behavior, setting, or broken rules (see \cite{Busselle2008}). Posts or comments described “odd jumps,” abrupt transitions, or storylines that collapse into “slop logic”—such as clips where characters instantly acquired money, switched locations without explanation, or performed actions unrelated to previous frames. Some highlight incoherent casts of characters: “First there was wood man then coffee girl and now troll man. The AI slop cinematic universe is getting crowded.” Others highlight characters that act incoherently: Death isn’t permanent and males getting pregnant. Some posters suggest this slop logic has “ruined [their] attention span.”

\textbf{Factual distortions} include instances where users point to outright falsehoods or fabricated realities embedded in slop-related discussions. Users described a range of fabrications—from invented celebrity deaths to staged political endorsements to fake animal footage, ahistorical imagery, and non-existent species—circulating as if they were genuine. The proliferation of AI-generated historical content was critiqued as making it “basically impossible to find decent content with ACTUAL history.” Other concerns involved deceptive medical advice and fear-based misinformation. One poster recounted how their parents abandoned prescribed treatments after watching AI-generated ads, adopted conspiratorial beliefs about foods being poisoned, and fell victim to multiple scams. 

In another vein, \textbf{provenance deficiencies} described content rejected for lacking authentic human creativity and intentionality, beyond assessments of its observable deficiencies. Users framed the defect not in terms of what the content depicted, but how it came into being. One poster articulated this distinction, stating “AI slop is slop wherever it’s sourced.” Similarly, when encountering an AI-generated image with a positive message about trans child acceptance, responses reflected this origin-based rejection: “Good message bad execution,” and “Is there a sub for redrawing AI slop? Because I feel this concept deserves to be drawn by a human artist,” indicating beliefs that value still requires authentic human execution to be legitimate. Content was labeled “slop” precisely as its genesis or sourcing was nonhuman—seen as automatic, effortless, generic, and devoid of human intention or creative labor. Even when acknowledging comparable quality issues in human-created content, posters maintained this categorical distinction; as one poster noted, “I can’t get over the fact you just implied Elsagate\footnote{A controversy of child-inappropriate content on YouTube.}  had ‘artistry and effort’ and, compared to AI slop, I had to agree,” conceding that even notoriously low-quality human content possessed something AI-generated material fundamentally lacked. The defect centered on origin: Content was rejected not for being visually wrong or morally disturbing, but simply because it was AI rather than human.

\subsection{Algorithmic Overreach}
Algorithmic overreach was the second major theme, capturing user frustrations, exhaustion, distrust, disappointment, or even outrage directed not at content itself, but at the platforms behind the algorithms that continually surface and circulate slop. 

The most common subtheme was \textbf{content saturation}, describing their feeds as increasingly cluttered, repetitive, and flooded. Posters discussed deluges of recycled AI-generated content produced by high-volume automated channels (often referred to as “content farms”). The monotonous repetition of slop was a persistent complaint. Posters observed identical tropes recurring on their feeds and eventually “wound up into collections of one single video.” They noted seeing the same AI-powered plush-toy ads, the same cat protagonist across slightly altered plotlines, and videos circulating under recurring, unrelated hashtags (e.g., \#jaidmeel) with millions of views, prompting one poster to observe: “There really is not a single original idea in the AI slop ecosystem. Is there?” Posters characterized this overwhelming influx as a “tsunami” flooding their social media platforms.

Slop was suggested to contribute to \textbf{platform decay}—the perceived deterioration of platform quality over time, attributed in part to platform priorities privileging monetization over user experience. Posts in this subtheme emphasized that platforms were “getting worse,” “slowly killing their reputation,” and “destroyed,” with some users describing shifts from “good in the past” to “a pit of hell now.” Posters pointed to declining recommendation relevance (“The search is BROKEN”), weakened governance (a slop article “redirected me to a malicious scam site.”), and increasingly opaque or coercive systems (“It is as if YouTube wanted to force me”) as evidence of systemic decline. Others interpreted these changes as the result of profit-driven design logics, arguing that platforms were willing to “make it worse” when financially advantageous. These accounts altogether framed decay as a trajectory of decline in which automation, monetization, and weakened oversight undermined user experience and rendered core platform features “monstrosities” within an ecosystem perceived as dominated by “bots” and “steaming heaps of trash.” 

Across discussions, another critique highlighted a perception of platforms prioritizing \textbf{automation over autonomy}, as posters described feeling a loss of agency as algorithmic processes steadily eclipsed personal choice, preference, and control. Posts suggested the problem with slop was not merely its existence, but the automated infrastructures that persistently propelled unwanted content into feeds, recommendations, and autoplay queues. Posters repeatedly emphasized the futility of controls, noting even after repeated attempts to hit “not interested” or “don’t recommend,” slop still resurfaced in their feeds. Posters also expressed frustration with platforms that disafforded the opting-out of slop—autoplay resumed itself, dislike functions were disabled or removed, and platform-level controls felt largely symbolic. “It is actually coded to send me that.... regardless of settings.” Together, these accounts characterized a broader interpretation of slop: a symptom of systems that prioritize automated distribution over individual autonomy, rather than a simply unwanted content.

Finally, posters discussed \textbf{engagement amplification}, pointing to suspicious visibility patterns in likes, comments, or views surrounding AI slop content that appeared designed to exploit algorithmic systems. They questioned, “How are AI slop posts like this one getting tens of thousands of likes. Is it botting?” Some noted bot-like and manipulative comment tactics designed to farm engagement and redirect viewers to AI slop channels. One poster discovered coordinated spam: “I found the same bot posting the same comment on 2 different videos gaining thousands of likes, which makes them appear at the top.” Posters also described comment sections that seemed disconnected from the video content itself, such as one where “the comments are just people spammed heart emojis” while the content was kittens in mortal danger of mutilation.  One poster described the phenomenon as: “Literal internet rot theory in action. Bots (are) reading AI slop, seeming to entangle “brain rot” from low-quality content and “dead internet theory” \cite{Muzumdar2025} to describe a feedback loop in which algorithms read bot-amplified engagement as human-user interest that, in turn, continues to circulate AI slop.

\subsection{User Resistance}
User resistance refers to the actions and interpretations posters used to push back against sociotechnical processes they believed were producing or amplifying slop. Posters described overt resistance in methods of refusal, critique, and reinterpretation that collectively challenged the conditions enabling slop’s visibility and circulation.

User resistance frequently took the form of \textbf{responsibility assignment}, through which posters sought to trace the chain of human actors responsible for creating and circulating slop. They drew attention to the human decisions, motives, and incentives that shaped slop’s production, as well as criticizing content creators, advertisers, distributors, and prominent tech leaders for making slop widespread, acting irresponsibly, or prioritizing profit over quality. Some comments explicitly targeted individuals. Some, for instance, remarked “Mark Zuckerberg is AI slop” and “He is AI slop before AI slop existed. Patient 0.” to critique him for having liked an AI-generated image on Facebook and perhaps hinting at the notion that he was inherently a precondition for slop. In this context, his engagement with such content was interpreted as emblematic of platform-leadership affinity for slop. Other posters focused on the labor practices and creative shortcuts that enable slop production. Posters criticized creators for investing minimal effort, asking, “Why would they put more effort into creating the prompt than any other element of the presentation?” Negative labels such as “sloppers,” “sloptubers,” or “grifters” were used to characterize slop creators and distributors, alongside critiques of advertisers for producing, appropriating, or reposting slop. They also argued that certain AI slop videos contained disguised fetish elements, often pointing to the repeated use of feline characters or other cute animals. According to users, animal figures sugarcoat low-quality (“crying cartoonishly”), nonsensical (“golden retrievers kneeling to pray”), or morally wrong material (“strawberry diaper cat on a crime spree”), making the underlying human motivations appear innocuous. 

\textbf{Agency reclamation} comprises a wide range of tactics aimed at reducing exposure to slop or disrupting the mechanisms that surface slop in feeds. Users described actions related to algorithmic avoidance (reporting, filtering, blocking, banning, erasing viewing history, requesting algorithm resets), platform refusal (quitting [“I am considering just switching to cable TV”], canceling subscriptions, downrating, migrating to alternative platforms), infrastructural interventions (signing petitions, sending feedback, protesting, mass reporting [“Enough! Redditors! Use the feedback function!... Share this, repost this, you need to use the Feedback Option or nothing will change. Get up your asses and be part of something bigger than yourself.”]). They also described content-level disruptions, such as leaving videos mid-watch in response to intrusive AI images—which may demote algorithmic priority. These tactics are often characterized as attempts to regain autonomy and control in an increasingly automated, opaque, and unsatisfying media environment. Yet the efficacy of these tactics remained uncertain. Many posters erased “a decade of viewing history” and clicked “not interested” hundreds of times, only to find slop persisting in their feed. 

\textbf{Societal risk articulation} involved critiquing slop as a broader threat that could endanger human existence or well-being. Users expressed concerns that slop could make people lose their jobs, crowd out genuine creators, undermine creative labor, and erode informational quality—concerns on par with those around image-generative AI more broadly \cite{Banks2025}. Others warned that slop was not age-restricted and impacted vulnerable groups such as older individuals or young audiences, “ruining the minds of our future generation.” Still others linked slop to larger, system-level harms, blaming it for “wrecking the economy” or exacerbating environmental degradation: “Imagine wasting several gallons of water to trick a bunch of grandmas. That’s depressing af.” “Trees die for this. Lakes dry for this. The atmosphere chokes for this.” 

\subsection{Entertainment Outcomes}
Entertainment outcomes, across discussions, were a less prominent but theoretically important motif that captured both hedonic and eudaimonic experiential outcomes \cite{Oliver2010} associated with AI slop. These included a range of hedonic experiences from pure fun, enjoyment, amusement, love, to a more complex, guilty pleasure that coalesced into a state of ambivalent enjoyment; those experiences presented alongside more indirectly eudaimonic entertainment outcomes linked to feelings of loss of, longing for, and appreciation of earlier forms of online culture. 

A form of entertainment outcome was \textbf{enjoyment}. Some posters reported they “ironically never skipped” any of them; others described them as funny, goofy, or even educational (“And finally something AI slop that teaches something”). Some posters found some AI-generated cats featured in these videos cute (“This is the AI slop that I want.”) and pointed out an example of Donald Trump swimming in a fish tank they believed was high-quality (“This is high-quality AI Slop”). Others went further and defended the content they liked by distancing it from an established definition of slop as low-quality content, arguing that what they enjoyed did not fit that label: “...AI should absolutely be used to create content that is so absurd and goofy that we can laugh...”. The label was further rejected by creators who found value in their own AI-generated content. One claimed we should not take hybrid content (that which combines AI-generated material with deliberate and intensive creative human labor) and collapse into slop because doing so erases all the human effort behind it: “AI is just a tool among others. Why is this one disqualifying by default?” Yet, these expressions of valued enjoyment often coexisted with users’ acknowledgment of the content’s absurdity and disturbance. For instance, one responded to a video depicting people falling from a cliffside thrill ride: it was their “favorite brand of AI slop where it is this fucking insane disaster.” This tension between pleasure and disapproval shaped much of the hedonic appeal of slop.

Another entertainment outcome centered on \textbf{technostalgia}—the yearning for rituals, practices, or ways of life associated with past technologies and online culture \cite{Niemeyer2023}. Nostalgia extended beyond specific technologies to the broader online culture and platform experiences they linked with pre-AI internet. Some expressed feelings of loss of and longing for an earlier era when they believed “the internet was decent,” when content felt more genuine, or when YouTube was “old YouTube.” The old YouTube was characterized by human creativity, personal expression, and senses of authenticity. It was a time when “people post[ed] a video because they were doing something” rather than producing content to target algorithms, when platform features like annotations and customizable channel pages allowed creators to make their channels “feel personal,” and when favorite channels produced content that users genuinely “enjoyed.” Slop, for many, marked the end of this era. One poster mourned the transformation of channels they had grown up with (“Our childhood just got turned into AI slop”). These rather eudaimonic responses framed AI slop not as merely a contemporary annoyance, but as a symbol of perceived decline in media culture, prompting users to reflect on how entertainment has changed over time, and what they value in entertainment experiences. Importantly, this was not a eudaimonic effect in the usual sense (i.e., media content prompting appreciation for the subject of that content; \cite{Oliver2011}), but instead slop appears to function as a trigger for reflecting on the old, anti-slop internet—and appreciating that lost internet. 

\begin{table*}[t]
\centering
\footnotesize
\renewcommand{\arraystretch}{1.35} 
\setlength{\tabcolsep}{6pt} 
\caption{Summary Table of Thematic Hierarchy on AI Slop Online Discussions}
\label{tab:slop_thematics}
\vspace{0.7em}
\begin{tabular}{|p{3.2cm}|p{3.8cm}|p{10.2cm}|}
\hline
\textbf{Theme} & \textbf{Subtheme} & \textbf{Conceptually relevant codes} \\
\hline
\multirow{5}{*}{\parbox{3cm}{Constitutional defects \\ (\textit{n} = 305)}}
    & Visual degradation \newline (\textit{n} = 108)
        & Melting or fused objects, merged body parts (creature hybrid), distorted faces, face blending, stitch together, lazy overlay, misspellings\\
\cline{2-3}
    & Moral transgression \newline (\textit{n} = 92)
        & Animal in distress, child in distress, implied murder, animal abuse, child abuse, bullying, rape, transphobia, blasphemy, sexism, arson, cheating, violence, racism, child abduction, bank robbery, stealing, frame, culture-war bait, pro-monarchy, trans-exclusionary radical feminist, islamophobic, anti-Muslim, destruction, fear induction, emotional exploitation\\
\cline{2-3}
    & Narrative inconsistency \newline (\textit{n} = 63)
        & Unexpected turns, odd jumps, incoherence, surrealism, incomprehensible, slop logic (instant money), unnecessary, unrelated\\
\cline{2-3}
    & Factual distortion \newline (\textit{n} = 39)
        & Scam, spam, fake news, ahistorical, misinformation, disinformation, fake ads, trick, inaccurate, digital falsification, twisted truths\\
\cline{2-3}
    & Provenance deficiencies \newline (\textit{n} = 35)
        & AI-generated, not human-made, creator degradation, no originality, generic, does not feel genuine, auto-generated in the slightest, uncanny\\
\hline

\multirow{4}{*}{\parbox{3cm}{Algorithmic overreach \\ (\textit{n} = 110)}}
    & Content saturation \newline (\textit{n} = 43)
        & Feed cluttering, ad-revenue farming, tsunami, content farm, boring, repetitive, same, again, recycled, certain hashtags (\#jaidmeel)\\
\cline{2-3}
    & Platform decay \newline (\textit{n} = 41)
        & Predator, irrelevant recommendations, bot-amplified, bypass algorithm, monstrosity of a feature, steaming heaps of trash, cooked, worse platform\\
\cline{2-3}
    & Automation over autonomy \newline (\textit{n} = 30)
        & Cannot opt out, autoplay, downvote, dislike disabled by owner/platform, loss of control, profit comes above user enjoyment, feels super forced\\
\cline{2-3}
    & Engagement amplification \newline (\textit{n} = 19)
        & Bot-amplified, algorithm-driven boosting, disproportionate visibility, suspicious engagement spikes, disproportionate engagement, likes boosting, engagement farming\\
\hline

\multirow{3}{*}{\parbox{3cm}{User resistance \\ (\textit{n} = 103)}}
    & Responsibility assignment \newline (\textit{n} = 60)
        & Human slop, Mark Zuckerberg, Sam Altman, creator, sloppers, sloptuber, grifters, irresponsible people massively sharing it, advertiser, stolen and reedited, disguised fetish, sugarcoat\\
\cline{2-3}
    & Societal risk articulation \newline (\textit{n} = 25)
        & Threats to human existence, not age-restricted, crowd out genuine creators, impacting elderly, ruining the minds of our future generation, wrecking economy, environment exacerbation\\
\cline{2-3}
    & Agency reclamation \newline (\textit{n} = 24)
        & Avoid, boycott, filter, report (AI police), sign the petition, ban, shut off, quit, bring back old one, protest, migration to alternative platforms, third-party tools, send feedback, cancel subscription, low rating, block, ask for algorithm reset, switch to cable TV, left a video mid-watch, do not make AI slop, erase viewing history to escape, mocking/satire\\
\hline

\multirow{2}{*}{\parbox{3cm}{Entertainment outcomes \\ (\textit{n} = 80)}}
    & Enjoyment \newline (\textit{n} = 68)
        & Entertaining, love, amusing, guilty pleasure, unironically enjoy, ironically never skip an AI slop video, adorable, wholesome, laugh, educational (teach something), cute, favorite AI slop story, benefit, goofy, W, high quality, attempt to distance AI-made content from AI slop\\
\cline{2-3}
    & Technostalgia \newline (\textit{n} = 12)
        & Dead internet, when the internet was decent, go back to ads for books and music, when I was a kid, miss the old Youtube, bring back the old Youtube\\
\hline
\end{tabular}
\end{table*}

\section{Discussion}
This study offered a descriptive account of how everyday users conceptualize AI slop. Results indicate folks frame AI slop as a multidimensional sociotechnical phenomenon organized around four themes: constitutional defects, algorithmic overreach, user resistance, and entertainment outcomes. Across these themes, the nature of slop is relatively cohesive (defective, invasive content) but value-laden reactions to them are mixed and even contested (resistance and loss versus enjoyment and creativity). Constitutional defects are legible via visual, textual, logical, moral, and narrative degradations as they disrupted coherence, truth, intelligibility, and expected or desired sourcing from human creators \cite{Millet2023}. In algorithmic overreach,  platforms themselves become slop farms, saturating feeds with whatever captures our attention most efficiently; our perceptual systems are more attuned to threats to content representing existential threat, depicting moral erosion, fostering emotionally alarm, and conveying visual chaos \cite{Brady2020, Stolte2021}. User resistance manifests as burdensome, often-futile acts against algorithms and their creators, mostly through reporting, disengagement, or exit. Varied entertainment outcomes complicate portrayals of slop as purely problematic: Slop is also genuinely enjoyed, a guilty pleasure, a technostalgia trigger.

In all, what counts as slop is negotiated at these intersections, so a definition of slop grounded in user experience must therefore extend beyond “low quality” alone to account for how such content is circulated, contested, and sometimes valued.	

\subsection{Toward Conceptual and Operational Definitions of AI Slop}
\begin{figure*}[t]
    \centering
    \includegraphics[width=\textwidth]{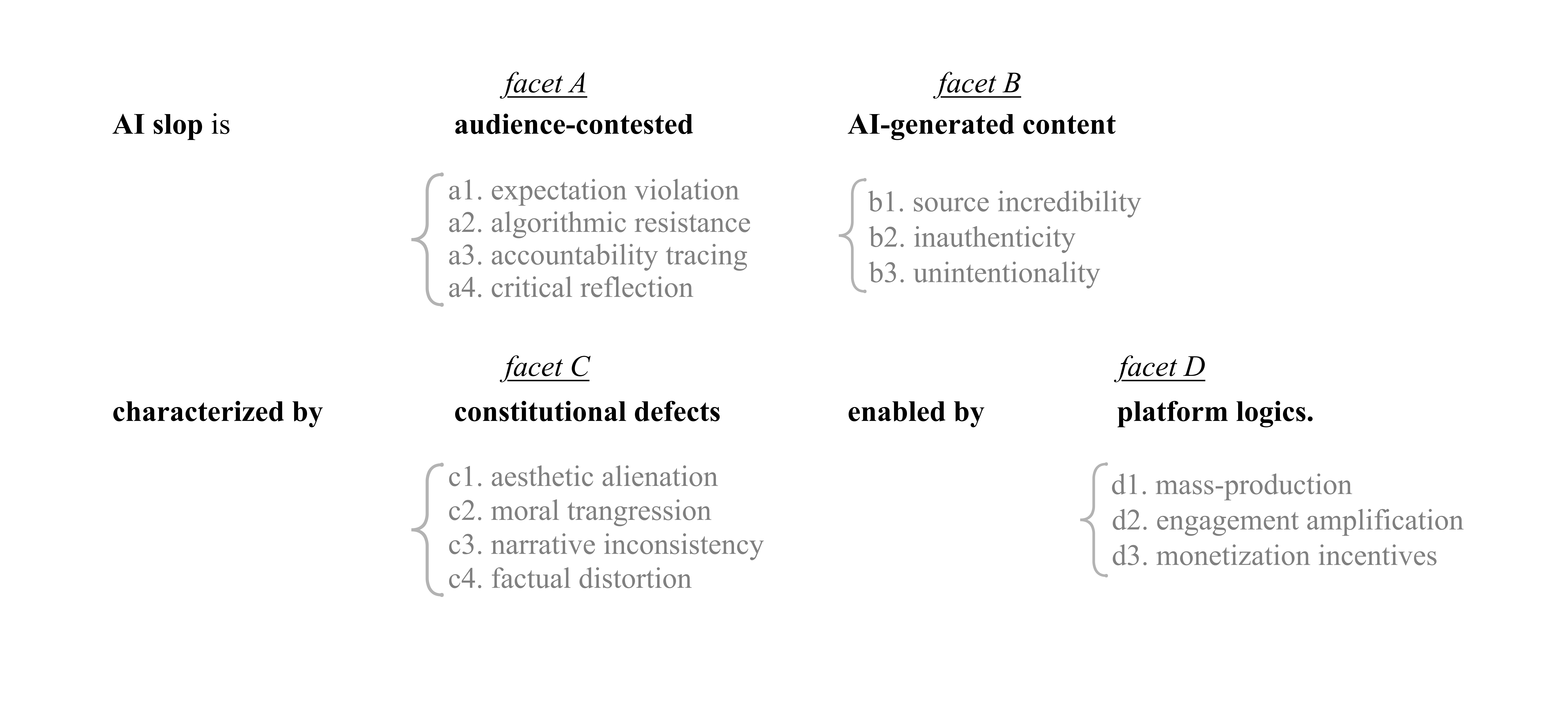}
    \caption{Mapping sentence defining the concept of AI slop.}
    \label{fig:mapping}
\end{figure*}

Informed by the interconnected themes above, we propose the following working definition for visual AI slop: \textbf{AI slop is audience-contested, AI-generated content characterized by constitutional defects and enabled by platform logics}. We further operationalize this definition through a mapping sentence (Figure \ref{fig:mapping}), which is a structural and textual framework that semantically maps out important variables and their associations \cite{Guttman1998, Hackett2021}.

In this conceptualization, audiences are active sense-makers who bring normative expectations of credibility, authenticity, and intentionality to slop encounters. Contestation (facet A) is multidimensional—the same content may provoke critical resistance and/or reflections in one context and ambivalent enjoyment in another, depending on valence and exposure. As AI-generated content (facet B), it carries questions of source credibility, authenticity, and intentionality—its provenance is machine-made, its authorship diffuse, and its intent ambiguous. The constitutional defects (facet C) are not merely aesthetic failures but sociotechnical ones, destabilizing the shared norms of accuracy, morality, and consistency that ground participatory culture. Platform logics (facet D) are not incidental but structurally generative—slop does not merely circulate through platforms, it is produced as a function of monetization, automation, and engagement metrics, platforms prioritize. The audience response is therefore not a reaction to content alone but to sociotechnical dynamics of production, circulation, and reception. AI slop is therefore best understood not as a fixed aesthetic category but as a \textit{fuzzy set} \cite{Zadeh1965} of relational and affective artifacts, defined by the gap between what algorithmic systems produce at scale and what audiences expect, contest, and occasionally embrace.

The mapping sentence scaffolds future inquiries. By combining elements across facets, researchers can generate structured configurations animating specific research questions. For instance, combining expectancy violation (a1), unintentionality (b2), aesthetic alienation (c1), and mass-production (d1) invites questions about whether AI content gradually reshapes audiences’ aesthetic expectations—and whether what once registered as uncanny becomes unremarkable over time. Combining accountability tracing (a3), unintentionality (b3), moral transgression (c2), and monetization incentives (d3) opens questions about how moral responsibility gets reassigned in AI slop ecosystems—users tend to blame human actors such as creators and platforms rather than AI systems, even as the transgressions are unintentional and platform profit logics obscure the actual chain of causation. How do audiences construct accountability narratives when the agents most responsible are least visible? In combination with relevant theory, such facet combinations can generate a systematic and theoretically grounded research agenda for understanding the operation and experience of slop.

\subsection{Considering Sloppy Tensions}
With this conceptualization in place, we turn to what user encounters with slop reveal about broader tensions in platformed media environments. We organize the following discussion around three potential stances that emerged from the data: AI slop as occupying an uncanny aesthetic and provenance space, AI slop as evasive and masquerading, and AI slop as fast food. Each stance carries distinct practical implications, which we develop in turn. 

\subsubsection{\textbf{AI Slop Occupies an Uncanny Space of Aesthetics and Provenance}}
AI slop introduces a form of content ambiguity that existing aesthetic and provenance frameworks are ill-equipped to handle. Its ambiguity operates across two dimensions simultaneously. Aesthetically, AI slop sometimes occupies an uncanny middle ground—neither convincingly hyper-realistic nor obviously synthetic, defying easy categorization \cite{Smith2026}—humanoid animals, hybrid creatures, and plastic-look visuals that trigger recognition without resolving into familiar categories. On provenance, users are sometimes left uncertain about the origin, authorship, or degree of human labor behind images—whether a human meaningfully shaped the output or whether it was generated instantaneously from a minimal prompt. The uncanny valley \cite{Mori2012} offers a partial account of the first dimension, theorizing the unease produced by near-human appearance, but was not designed to handle content whose ambiguity is as much about origin as aesthetics. We extend it to capture what we call the \textit{uncanny of provenance}, that is disorientation arising not just from how content looks, but from the irresolvability of who made it. The dual ambiguity was visible in our data where posters developed folk taxonomies to distinguish AI slop from creative AI-generated content, reflecting attempts to resolve provenance uncertainty through informal heuristics. Yet, these distinctions were difficult to sustain, as aesthetic markers once indexing human labor are now reproducible at scale without it. This \textit{uncanny of provenance} has direct consequences for how platforms govern and label AI-generated content. 

Platform-specific labeling practices (i.e., whether content is AI generated, or not) are deployed under the banner of transparency. However, our data indicate users do not engage the labels uniformly, such that labels could be they actually undermine users’ epistemic agency—their capacity to make informed judgments about the content they encounter \cite{Coeckelbergh2023}—in relation to the information on which they actually base judgments. Current platform binary labeling practices for AI content, partnered with the C2PA—including TikTok’s disclosure requirement for “AI-generated,” YouTube’s “altered or synthetic content” tag, and Meta’s “AI info” label—give them insufficient signals to act on user intuitions and values. These frameworks share a common foundational logic: disclosure is triggered only when content appears realistic enough to potentially deceive viewers. Content that is obviously synthetic or stylized falls outside mandatory requirements (“Clearly unrealistic content, such as animation or someone riding a unicorn through a fantastical world;” \cite{YouTube2024}), leaves voluntary disclosure or automated provenance detection to fill the gap. As a result, these labels, while not without value, fall short of the nuanced information users draw on when evaluating content \cite{Fisher2018, Moruzzi2025, Wittenberg2025}. 

\subsubsection{\textbf{AI Slop Evades and Masquerades}}
Reading across induced themes, AI slop’s spread was seen as facilitated by two interlocking exploitations of the structural limits of platform moderation. The first is evasion. Posters described otherwise unrelated YouTube Shorts as repeatedly carrying the same hashtag despite no connection among them. We interpret this pattern to suggest that viral, non-AI specific hashtags may embed slop within broader streams of human-generated content, potentially complicating users’ keyword- and description-based filtering \cite{Vera2025} and extending its visibility through platform recommendation systems. The second is masquerading—the adoption of aesthetic softening, wherein videos take on cutesy, cartoonish, or affectively appealing qualities that allow them to pass both the “deceiving hyper-realistic” threshold required for labeling and to disarm viewer wariness. Together, these tactics exploit the structural gap between what moderation systems are designed to catch—explicit rule violations and formal compliance \cite{Gerrard2018, Gorwa2020}—and the actual qualities that define slop for users.

Slop’s persistence, then, may not reflect an oversight in moderation design so much as predictable outcome of moderation systems built around explicit rule violations while the content in question may circulate without triggering established thresholds for intervention \cite{Gerrard2018,Weener2026}. A hashtag such as “\#cutecats” violates no policy: It is out of scope for YouTube’s hashtag policy prohibiting hate speech, sexual content, vulgar language \cite{YouTubeHelp}. And a cartoonish humanoid animal violates no policy (for example, TikTok’s AI-generated content policy on hyper-realism; \cite{TikTok2026a}. The result is a systematic referral of interpretive and resistive labor onto users, who are left to identify and counter unwanted content that platforms have structurally failed to address. This misalignment becomes particularly clear when we consider what platform features like “dislike” or “not interested” are actually designed to do, as preference calibration devices allowing users to signal disinterest and shape subsequent recommendations \cite{Liu2024}, rather than provide users with direct means of excluding slop as a category. It remains unclear, though, whether these affordances can be repurposed to carry the interpretive weight users currently shoulder on their own. That reconfiguration may require a new understanding of what these (dis)engagement features actually afford \cite{Molina2022,Scharlach2023}, and what they would need to afford, if they meaningfully register for contestation rather than mere preference.

\subsubsection{\textbf{AI Slop is Fast Food}}
AI slop is not just difficult to recognize or resist—it is a symptom of a production logic oriented toward volume, speed, and low cost over quality or authenticity, mirroring what Ritzer’s notion of “McDonaldization” \cite{Ritzer2021} posited: the rationalization of production around efficiency (fastest path to output), calculability (quantity over quality), and predictability (replicable templates over original craft). Much like fast food, slop is engineered for rapid consumption, volume production, low-cost replicability. It is palatable on the surface but often falls short of what it presents itself to be. In this way, slop is an unsurprising output of incentive structures that reward high-volume posting and broad reach over depth, originality, or creativity. Posters in our data did not simply describe slop as a bad content—it felt engineered to be consumed without resistance, delivering rapid stimulation and affective reward much as fast food is formulated to trigger dopamine-driving craving cycles \cite{Morales2020}. Others directed blame not just at platforms for setting the conditions but at individual slop creators who opted into that production logic. In this sense, everyday users echo sentiments around “the irrationality of rationality” as systems are optimized for efficiency that end up dehumanizing, stripping workers and consumers alike of autonomy and creativity, and ultimately undermining the conditions of their own value (\cite{Ritzer2021}, p. 378). 

Implications for platform design are non-trivial. Data suggest users would welcome a recalibration of creator monetization programs that offers some relief from unwanted slop. TikTok’s Creator Rewards Program \cite{TikTok2026b}, for instance, increasingly rewards quick interactions, rapid growth, and high-volume posting. This creates a “regime of virality” \cite{Lee2026} where platform monetization structurally forces creators to mass-produce cheap, rapid content just to survive the algorithm. Even as its own eligibility criteria describe “reward-worthy content” as “original,” “well-crafted,” demonstrating “unique perspective”—criteria grounded in forms of creativity understood as distinctly human \cite{Runco2023}. But one must ask: Can a prompt-based creation process constitute “well-crafted” content? Can mass-produced AI output demonstrate “unique perspective?” These answers remain absent from platform policy, leaving a structural gap that everyday users recognize. Indeed, some labeled the phenomenon “human slop” to conceptualize the problem as rising not only in AI systems but in human decisions and incentives that sustain them.

\subsection{Limitations}
Our study is subject to the usual limitations of the method. We gathered content from online forums arguably frequented by people who are likely to care enough about the topic to discuss it with others (a purposeful decision to examine social constructions of the focal concept). The data represent a snapshot of sentiments in the bounded sampling time frame (carefully selected to span the emergence of the phenomenon) when technologies and attitudes are likely to change over time. The analysis was conducted by one analyst with a particular lens (mitigated by the critical review by the second author). Additionally, although “AI slop” has recently gained visibility in mainstream media and dictionary definitions, the term itself may not be uniformly used by social media users. Many users might describe the same phenomenon using alternative expressions such as “low-quality AI-generated content,” “AI spam,” “AI brainrot,” or simply “slop” or “AI-generated” without adopting the specific label “AI slop.” As a result, the data collection strategy might overrepresent users who were familiar with or explicitly engage with this particular term, potentially reflecting higher levels of digital literacy or participation in niche online discourses \cite{Lutz2022}. As a result, some concerns identified in this study, such as narrative inconsistency or visual degradation, may not persist as AI generation tools (and literacy around them) continue to improve.

\subsection{Future Directions}
Our findings and interpretations animate several future research directions. Perhaps most fundamentally, AI slop—as a fuzzy-set, contested category of generative content sustained by platform logic—invites a destabilization of traditional categories humans use to evaluate media: quality, authenticity, meaning, effort, and creativity. Future research might ask whether we need new epistemological taxonomies in these shifts, and our findings gesture toward at least three concrete directions.

Our conceptual definition (Figure \ref{fig:mapping}) surfaces a set of research questions about the \textit{values and expectations} that underlie reactions to slop. When users label content as slop, what aesthetic, moral, and relational standards are they invoking, and how do these standards trade off against one another? A user who is aesthetically forgiving (tolerating text that dissolves into gibberish or fusions between humans and fruits) may nonetheless be morally strict about content that exploits real tragedy, fabricates suffering for engagement, while others may weigh these dimensions in reverse such that there are idiosyncratic hierarchies. Future work could test this by systematically varying defect type to identify which violations most reliably trigger the slop label, converting our thematic typology of constitutional defects into a testable hierarchy distinguishing non-compensatory (disqualifying) from compensatory cues \cite{Billings1983}. Such a hierarchy would carry direct implications for AI content generation. If certain defects (e.g., moral transgression) function as disqualifying regardless of surrounding quality, while others (e.g., visual degradation) are accepted when offset by narrative strengths, generation systems could be tuned to avoid former even at the cost of the latter. Perhaps we can also imagine a different future for this line of inquiry that moves beyond optimizing away disqualifying cues, toward how human- and AI-generated content can co-exist—where both the provoked contestation and appreciation become windows into ongoing, resolved processes of human meaning-seeking.

This path of meaning-seeking, however, is shaped by the platforms on which contestation and appreciation of AI-generated content take place. A second direction turns to \textit{platform logics}. To what extent is slop a property of content itself, and to what extent is perception of slop shaped by revenue models, creator incentives, and governance structures through which content circulates? How might these platform logics shape whether users reject, tolerate, contest, or find value in what they perceive as slop? The gap between users’ agency and the affordances available to them may be structural, since engagement-based business models (e.g., TikTok, Instagram) can derive value from the very content users contest \cite{Liang2022, Mochon2024}. Users may respond differently to the same or similar forms of “slop,” rejecting or contesting them in some cases while finding entertainment or other value in others—yet both positive and negative responses can generate engagement through sharing, reposting, argument. This raises a further question: Does slop become more visible or more persistent where engagement is the primary currency, particularly when both resistance and enjoyment feed the algorithms that sustain its circulation?  If it is correct, “slop” may not behave consistently across platforms with different revenue models, creator incentives, and governance structures, as these differences can shape how content is moderated, made visible, and circulated \cite{Bono2024, Liu2022, Meng2026}. 

Importantly, platforms are not the only force driving the boundaries of slop. A third direction sits at the nexus of users, platforms, and regulators, querying: how do they each \textit{differently define what counts as slop}. AI slop may not be defined uniformly across the actors who encounter or theorize it—including in comparison to our own user-driven definition above. Data analyzed here suggest users orient toward authenticity and creativity, and they see platforms orienting toward profitability and efficiency. In tandem, regulators may prioritize disclosure and harm reduction (e.g., \cite{CATransparencyAct2024}), and researchers may lean toward analytic soundness and generalizability \cite{Polit2010}. Future research should examine whether these conceptualizations or operationalizations converge or diverge across contexts and with what consequence. Altogether, these questions addressed could reveal not a static or constrained boundary but an ongoing negotiation over whose definition and experience prevails. 

\section*{Data Availability Statement}
Data gathered and analyzed in this study comprise raw forum posts and are not shared to avoid search-driven identification of posters. A paraphrased version of the dataset, along with the data collection script and coding procedure, is available at \href{https://osf.io/xzcjs/overview?view_only=6313338f9e0a4e0dbbe034ac4fcf09a3}{https://osf.io/xzcjs}.

\begin{acks}
Z.L. would like to thank Dr. Jennifer Stromer-Galley for her helpful notes on an early version of this work.
\end{acks}

\bibliographystyle{ACM-Reference-Format}
\bibliography{MAIN}

\appendix


\end{document}